\documentclass[11pt,epsfig]{article}
\usepackage{graphicx}

 1
\font\elevenrm=cmr10 scaled\magstep 1

\renewenvironment{thebibliography}[1]
 { \elevenrm
   \begin{list}{\arabic{enumi}.}
    {\usecounter{enumi}     \setlength{\parsep}{0pt}
     \setlength{\itemsep}{3pt} \settowidth{\labelwidth}{#1.}
     \sloppy
    }}{\end{list}}

\begin{document}

\title{{\bf Nuclear glory phenomenon}   \\
('Buddha's light' of cumulative particles) \footnote{The talk presented at the 18th International Seminar on High Energy Physics (Quarks 2014)	
2-8 Jun 2014. Suzdal, Russia, CNUM: C14-06-02.6;
11th Conference Quark Confinement and Hadron Spectrum, 7-12 September 2014, St Petersburg, Russia; 
Session of the Nuclear Physics Department of RAS, 17 - 21 November 2014, MEPHI, Moscow.
Based on the papers \cite{kmp,long}.}}
\author{V.~B.~Kopeliovich$^{a,b}$\footnote{{\bf e-mail}: kopelio@inr.ru},
G.~K.~Matushko$^{a}$\footnote{{\bf e-mail}: matushko@inr.ru},
and I.~K.~Potashnikova$^c$\footnote{{\bf e-mail}: irina.potashnikova@usm.cl}
\\
\small{\em a) Institute for Nuclear Research of RAS, Moscow 117312, Russia}
\\
\small{\em b) Moscow Institute of Physics and Technology (MIPT), Dolgoprudny, Moscow district, Russia} 
\\
\small{\em c) Departamento de F\'{\i}sica, Universidad T\'ecnica Federico Santa Mar\'{\i}a;}\\
\small{\it and Centro Cient\'ifico-Tecnol\'ogico de Valpara\'iso,
Avda. Espa\~na, 1680, Valpara\'iso, Chile}
}

\date{}
\maketitle

\begin{abstract}
Analytical explanation of the nuclear glory effect, which is similar to the known optical (atmospheric)
glory phenomenon, is presented. It is based on the small phase space method for the multiple
interaction processes probability estimates and leads to the characteristic angular dependence 
of the production cross section $d\sigma \sim 1/ \sqrt {\pi - \theta}$ in the vicinity of 
the strictly backward direction, for any number of interactions $N\geq 3$, either elastic or 
inelastic. Rigorous proof of this effect is given for the case of the optimal kinematics, as 
well as for arbitrary polar scattering angles in the case of the light particle rescattering,
but the arguments in favor of the backward azimuthal (axial) focusing are quite general and hold 
for any kind of the multiple interaction processes. Such behaviour of the cross section near
the backward direction agrees qualitatively with available data.
In the small interval of final angles including the value $\theta =\pi$ the angular dependence of the
cumulative particle production cross section can have the crater-like (or funnel-like) form. Further studies
including, probably, certain numerical calculations, are necessary to clear up this point.
\end{abstract}

\section{Introduction} 
The studies of the particles production processes in high energy interactions
of different projectiles with nuclei, in regions forbidden by kinematics for the 
interaction with a single free nucleon, began back in the  70th mostly at JINR (Dubna), headed by A.M.Baldin and
V.S.Stavinsky, and at ITEP (Moscow) where during many years the leader and great enthusiast of these studies 
was professor G.A.Leksin.
 Relatively simple experiments could provide 
information about such objects as fluctuations of the nucleus density \cite{blokh} or, discussed much
later, few nucleon (or multiquark) clusters probably
existing in nuclei. At JINR such processes have been called "cumulative production"
\cite{baldin1,baldin2}, at ITEP the variety of properties of such reactions has been called
"nuclear scaling" \cite{leksin1}- \cite{leksin3} because certain universality of these properties
has been noted, confirmed somewhat later at much higher energy,
$400\, GeV$ incident protons \cite{fran-le-1} and for $40\,GeV/c$ incident pions, kaons and antiprotons
\cite{antip1}.
A new wave of interest to this exciting topic appeared lately. New experiment has been performed
in ITEP \cite{abramov} aimed to define the weight of multiquark configurations in the carbon
nucleus. 
The interpretation of these phenomena as being manifestation of internal structure of
nuclei assumes that the secondary interactions, or, more generally, multiple
interactions processes (MIP) do not play a crucial role in such reactions \cite{konshma} - \cite{galoian}. 

The development of the Glauber theory \cite{glauber} to the description of particles scattering off
nuclei has been considered many years ago as remarkable progress in understanding
the particles-nuclei interactions. Within the Glauber model the amplitude of the                                 
particle-nucleus scattering is presented in terms of elementary particle-nucleons
amplitudes and the nucleus wave function describing the nucleons distribution inside
the nucleus. The Glauber screening correction for the total cross section of
 particle scattering
off deuteron allows widely accepted, remarkably simple and transparent interpretation.

Gribov \cite{gribov} investigated nontrivial peculiarities of the space-time picture of such scattering
processes and concluded that the inelastic shadowing corrections play an important role at
high enough energy and should be included into consideration. 
In the case of the large angle particle production the background processes which mask 
the possible manifestations of nontrivial features of nuclear structure, are 
subsequent multiple interactions with nucleons inside the nucleus leading to the
particles emission in the "kinematically forbidden" region.
Leonid Kondratyuk first noted that rescattering
of intermediate particles could lead to the final particles emission in
"kinematically forbidden" regions (KFR).
The double interaction process in the case of the pion
production off deuteron has been investigated first in
\cite{konkop}. Later the multiple interaction processes leading to nucleons
production in KFR were studied in \cite{kop1}
where the magnitude of the cumulative protons production cross sections was estimated as well.

 M.A.Braun and V.V.Vechernin with coauthors made some important observations concerning
processes leading to the particles emission in KFR \cite{bv-res-1}-\cite{braun-lepto}, including the processes 
with resonances in intermediate state \cite{bv-res-1}.
They found also that processes with pions in intermediate state
lead to the nucleons emission in KFR due to subsequent processes, like $\pi\,N \to N\,\pi$ \cite{braun-pin1}.
Basic theoretical aspects of MIP leading to the cumulative particles emission and some review of the situation 
in this field up to 1985 have been presented in \cite{long}.

Several authors attempted the cascade calculations of cumulative particles production
cross sections relying upon the available computing codes created previously 
\cite{sibste} - \cite{nomad1}.
The particles production cross section was found to be in moderate agreement with data.
Different kinds of subprocesses play a role in these calculations, and certain work
should be performed for detailed comparison. In calculations by NOMAD Collaboration the particles
formation time (length) has been considered as a parameter, and results near to the experimental
observations have been obtained for this time equal to $\sim 2 \,Fm$ \cite{nomad1}.

While many authors admitted the important role of the final state interactions (FSI), most of them 
did not discuss the active role of such interactions, i.e. their contribution to particles production 
in KFR, see e.g. \cite{size}.  Several specific features of the MIP mechanism have been noted experimentally and 
discussed theoretically \cite{kop1,kopmix, long},
among them the presence of the recoil nucleons, which amount grows with increasing energy of the
cumulative particle, possible large value of the cumulative baryons polarization, and some other, 
see \cite{long}. The enhancement of the production cross section near the strictly backward
direction has been detected in a number of experiments, first at JINR (Dubna) \cite{baldin-77, foc-1} and later at
ITEP (Moscow) \cite{foc-2, foc-4}. This glory-like effect
which can be called also the "Buddha's light" of cumulative particles, has been shortly discussed previously
in \cite{kop1, long}. More experimental evidence of this effect appeared since that time \cite{glo-1, glo-2}. 
We have shown analytically in \cite{kmp} that presence of the backward focusing  effect
is an intrinsic property of the multiple interaction mechanism leading to the cumulative particles production.
\section{Features of kinematics of the processes in KFR}
When the particle with 4-momentum $p_0=(E_0, \vec p_0)$ interacts with the nucleus with the mass $m_t \simeq A m_N$,
and the final particle of interest has the 4-momentum $k_f=(\omega_f, \vec k_f)$á
ôt large enough incident energy, $E_0 \gg m_t,\;M_f$, the restriction takes place
$$ \omega_f - z k_f \leq m_t, \eqno (2.1) $$
which is the basic restriction for such processes. $z=cos\,\theta < 0$ for particle produced
in backward hemisphere. The quantity $(\omega_f - z k_f)/m_N$ is called the cumulative number
(more precize, the integer part of this ratio plus one).

Let us recall some peculiarities of the multistep processes kinematics established
first in \cite{kop1} and described in details in \cite{long}. It is very selective 
kinematics, essentially different from the kinematics of the forward scattering off nuclei.

For light particles ($\pi$-meson, for example) iteration of the Compton formula
$${1\over \omega_{n} } -{1\over \omega_{n-1}} \simeq {1\over m} \left[1-cos (\theta_n)\right] \eqno (2.2)$$
allows to get the  final energy in the form                     
$$  {1\over \omega_N}- {1\over \omega_0} = {1\over m} \sum_{n=1}^N  \left[1-cos (\theta_n)\right] \eqno (2.3)  $$
The maximal energy of final particle is reached for the coplanar process when
all scattering processes take place in the same plane and each angle equals to
$\theta_k=\theta/N$.
As a result we obtain
$$  {1\over \omega_N^{max}}- {1\over \omega_0} = {N\over m}\left[1-cos (\theta/N)\right] \eqno (2.4) $$
Already at $N>2$ and for $\theta \leq \pi$ the $1/N$ expansion can be made (it is in fact the $1/N^2$ expansion):
$ 1-cos (\theta/N) \simeq \theta^2/2N^2 \left(1 - \theta^2/12N^2\right)$,
and for large enough incident energy $\omega_0$ we obtain
$$\omega_N^{max} \simeq N {2m\over \theta^2} + {m\over 6N}. \eqno (2.5)$$
This expression works quite well beginning with $N=2$.
Remarkably, that this rather simple property of rescattering processes has not been
even mentioned in the pioneer papers \cite{baldin1} - \cite{leksin3}
\footnote{This property was well known, however, to V.M.Lobashev, who observed experimentally
that the energy of the photon after 2-fold interaction can be substantially
greater than the energy of the photon emitted at the same angle in
1-fold interaction.}.

In the case of the nucleon-nucleon scattering (scattering of particles with equal 
nonzero masses in general case) the following approximate relation has been obtained for
the final particle momentum as a result of the  $1/N^2$ expansion 
at large enough $N$ and large incident energy 
$$k_N^{max} \simeq N {2m\over \theta^2} - {m \over 3N}, \eqno (2.6)$$
which coincides at large $N$ with previous result $(2.5)$ for the rescattering of light particles,
but preasymptotic corrections are negative in this case and twice greater. 
The normal Fermi motion of nucleons inside the nucleus makes these boundaries wider \cite{long}:
$$k_N^{max} \simeq N {2m\over \theta^2} \left[1 + {p_F^{max}\over 2m}\left(\theta + {1\over \theta} \right) \right], \eqno (2.7)$$
 where it is supposed that the final angle $\theta$ is large, $\theta \sim \pi$. For numerical estimates we
took the step function for the distribution in the Fermi momenta of nucleons inside
of nuclei, with $p_F^{max}/m\simeq 0.27$, see \cite{long} and references there. At large enough $N$
normal Fermi motion makes the kinematical boundaries for MIP wider by about $40$ \%.

There is characteristic decrease (down-fall) of the cumulative particle production cross
section due to simple rescatterings near the strictly backward direction. 
However, inelastic processes with excitations of intermediate
particles, i.e. with intermediate resonances, are able to fill up the region at $\theta \sim \pi$.

The elastic rescatterings themselves are only the "top of the iceberg".
Excitations of the rescattered particles, i.e. production of resonances in intermediate
states which go over again into detected particles in subsequent interactions,
provide the dominant contribution to the production cross section.
Simplest examples of such processes may be $NN \to NN^* \to NN$, $\pi N\to \rho N \to \pi N$,
etc. The important role of resonances excitations in intermediate states for cumulative
particles production has been noted first in \cite{bv-res-1}a and somewhat later in
\cite{kop1}. At incident energy about few GeV the dominant contribution into cumulative protons
emission provide the processes with $\Delta(1232)$ excitation and reabsorption, see \cite{long} and
\cite{dakhno}. Experimentally the role of dynamical excitations in cumulative nucleons production
at intermediate energies has been extablished in \cite{komarov} and, at higher energy, in \cite{malki}.
 
When the particles in intermediate states are slightly 
excited above their ground states, approximate estimates can be made. Such resonances could be $\Delta 
(1232)$ isobar, or $N^*(1470), \; N^*(1520)$ etc. 
for nucleons, two-pion state or $\rho(770)$, etc for incident pions, $K^*(880)$ for kaons.
This case has been investigated previously with the result for the relative change (increase)
of the final momentum $k_f$ (Eq. $(8)$ of \cite{kop1})
$$ {\Delta k_f \over k_f} \simeq {1\over N} \sum_{l=1}^{N-1} {\Delta M_l^2\over  k_l^2}, \quad
 {\Delta k_f^2} \simeq {2\over N^3} \sum_{l=1}^{N-1} l^2\Delta M_l^2, \eqno (2.8) $$
with $\Delta M_l^2 = M_l^2 - \mu^2$, $k_l$ is the value of 3-momentum in the $l$-th intermediate state.
This effect can be explained easily: the additional energy stored in the mass of intermediate particle
is transfered to the kinetic energy of the final (cumulative) particle.

\section{The small phase space method for the MIP probability calculations}
This method, most adequate for analytical and semi-analytical calculations of
the MIP probabilities, has been proposed in \cite{kop1} and developed later in 
\cite{long}. It is based on the fact that, according to established in \cite{kop1}
and presented in previous section kinematical relations, there is a preferable plane of the 
whole MIP leading to the production of energetic particle at large angle $\theta$, 
but not strictly backwards. Also, the angles of subsequent rescatterings are close
to $\theta / N$. Such kinematics has been called optimal, or basic kinematics.
The deviations of real angles from the optimal values are small, they are defined mostly
by the difference $k_N^{max} - k $, where $k_N^{max}(\theta)$ is the maximal possible momentum reachable
for definite MIP, and $k$ is the final momentum of the detected particle.
$k_N^{max}(\theta)$ should be calculated taking into account normal Fermi
motion of nucleons inside the nucleus, and also resonances excitation ---
deexcitation in the intermediate state. Some high power of the difference  $(k_N^{max} - k)/k_N^{max} $
enters the resulting probability.

Within the quasiclassical treatment adequate for our case, the probability product approximation is
valid, and the starting expression for the inclusive cross section of the particle
production at large angles contains the product of the elementary subprocesses matrix elements squared,
 see, e.g., Eq. (4.11) of \cite{long}.

After some evaluation, introducing differential cross sections of binary reactions $d\sigma_l/dt_l(s_l,t_l) $ instead of 
the matrix elements of binary reactions $M_l^2(s_l,t_l)$, we came to the formula for the production cross section 
due to the $N$-fold MIP \cite{kop1,long}
$$f_N(\vec p_0,\vec k)= \pi R_A^2 G_N(R_A,\theta) \int \frac{f_1(\vec p_0,\vec k_1) (k_1^0)^3 x_1^2dx_1 d\Omega_1}{\sigma_1^{leav}\omega_1}
\prod_{l=2}^N\left({d\sigma_l(s_l,t_l)\over dt_l}\right) \frac{(s_l-m^2-\mu_l^2)^2-4m^2\mu_l^2}{4\pi m\sigma_l^{leav} k_{l-1}} $$
$$\times \prod_{l=2}^{N-1}\frac{k_l^2 d\Omega_l}{k_l(m+\omega_{l-1}-z_l\omega_lk_{l-1})\;}{1\over \omega_N'}\delta(m+\omega_{N-1}-\omega_N-\omega_N').
\eqno(3.1) $$ 
Here $z_l= cos\, \theta_l$,
 $\sigma_l^{leav}$ is the cross section defining the removal (or leaving) of the rescattered object at the corresponding section
of the trajectory, it is smaller than corresponding total cross section. 
$G_N(R_A,\theta)$ is the geometrical factor which enters the probability of the $N$-fold multiple interaction with
definite trajectory of the interacting particles (resonances) inside the nucleus. This trajectory is defined mostly
by the final values of $\vec k$ $(k, \,\theta) $, according to the kinematical relations of previous section. 
Inclusive cross section of the rescattered particle production in the first interaction  is
$\omega_1 d^3\sigma_1/d^3k_1=f_1(\vec p_0,\vec k_1)$ and 
$d^3k_1=(k_1^0)^3x_1^2dx_1$,
$\omega_N=\omega$ --- the energy of the observed particle.

To estimate the value of the cross section $(3.1)$ one can extract the product of the cross sections out of the integral $(3.1)$ near the 
optimal kinematics and multiply by the small phase space avilable for the whole MIP under consideration \cite{kop1,long}.
Further details depend on the particular process. For the case of the light particle rescattering, $\pi$-meson for example, $\mu_l^2/m^2\ll 1$,
we have
$${1\over \omega_N'}\delta(m+\omega_{N-1}-\omega_N-\omega_N') = {1\over k k_{N-1}}
\delta\left[ {m\over k} - \sum_{l=2}^N (1-z_l) - {1\over x_1}\left({m\over p_0}+1-z_1\right)\right] \eqno(3.2) $$
To get this relation one should use the equality $\omega_N'=\sqrt{m^2+k^2+k_{N-1}^2-2k k_{N-1}z_N}$ for the recoil nucleon energy
and the well known rules for manipulations with the $\delta$-function.
When the final angle $\theta$ is considerably
different from $\pi$, there is a preferable plane near which the whole
multiple interaction process takes place, and only processes near this plane
contribute to the final output. At the angle $\theta =\pi$, strictly backwards,
there is azimuthal symmtry, and the processes from the whole interval of azimuthal 
angle $0< \phi < 2\pi$ provide contribution to the final output (azimuthal focusing, see next section).
A necessary step is to introduce azimuthal deviations from this optimal
kinematics, $\varphi_k$, $k=1,\,...,N-1$; 
 $\varphi_N=0$ by definition of the plane of the process, $(\vec p_0, \vec k)$.
Polar deviations from the basic values, $\theta/N$, are denoted as $\vartheta_k$, obviously, 
 $\sum_{k=1}^N\vartheta_k = 0$. The direction of the momentum $\vec k_l$ after $l$-th 
interaction, $\vec n_l$,  is defined by the azimuthal angle $\varphi_l$ and the polar angle 
$\theta_l = (l\theta /N) +\vartheta_1+...+\vartheta_l$, $\theta_N=\theta $.

Then we obtain making the expansion in $\varphi_l$,  $\vartheta_l$ up to quadratic terms
in these variables:
$$z_k= (\vec n_k \vec n_{k-1}) \simeq cos (\theta/N) (1-\vartheta_k^2/2) -sin (\theta/N) \vartheta_k 
+ sin (k\theta/N) sin [(k-1)\theta/N] (\varphi_k -\varphi_{k-1})^2/2. \eqno (3.3)$$
In the case of the rescattering of light particles the sum enters the phase space of the process
$$ \sum_{k=1}^N (1- cos\vartheta_k) =  N[1-cos(\theta/N)] + cos(\theta/N)
\sum_{k=1}^N\bigg[-\varphi_k^2\, sin^2(k\theta/N) + $$
$$+{\varphi_k \varphi_{k-1}\over cos(\theta/N)}sin(k\theta/N)sin((k-1)\theta/N)\bigg] -{cos(\theta/N)\over 2} \sum_{k=1}^N \vartheta_k^2 \eqno (3.4)$$ 
To derive this equality we used that $\varphi_N=\varphi_0=0$ --- by definition of the plane of the MIP,
and the mentioned relation $\sum_{k=1}^N\vartheta_k = 0$.
We used also the identity,  valid for $\varphi_N=\varphi_0 =0$:
$${1\over 2}\sum_{k=1}^N \left(\varphi_k^2+\varphi_{k-1}^2 \right) sin(k\theta/N)sin[(k-1)\theta/N] = 
cos(\theta/N)\sum_{k=1}^N\varphi_k^2 sin^2(k\theta/N). \eqno (3.5)$$
It is possible to present the quadratic form in angular variables which enters $(3.4)$ in the canonical form and to perform integration easily, see
Appendix B and Eq. $(4.23)$ of \cite{long}, and also \cite{kmp}.
As a result, we have the integral over angular variables of the following form:
$$ I_N(\Delta_N^{ext}) = \int \delta\biggl[\Delta_N^{ext} - z_N^\theta \bigg(\sum_{k=1}^{N} \varphi_k^2 -\varphi_k\varphi_{k-1}/z_N^\theta
+\vartheta_k^2/2\biggr)\biggr] \prod_{l=1}^{N-1} d\varphi_ld\vartheta_l 
= \frac{\left(\Delta_N^{ext}\right)^{N-2} (\sqrt 2 \pi)^{N-1}}{J_N(z_N^\theta) \sqrt N (N-2)! \left(z_N^\theta \right)^{N-1}}, \eqno (3.6) $$
$ z_N^\theta = cos(\theta/N)$. Since the element of a solid angle $d\Omega_l= sin(\theta\,l/N)d\vartheta_l d\varphi_l $, we made here substitution
$sin(\theta\,l/N)\,d\varphi_l \to d\varphi_l$ and $d\Omega_l \to d\vartheta_l d\varphi_l $,
$z_N^\theta = cos(\theta/N)$. 
The whole phase space is defined by the quantity
$$\Delta_N^{ext}\simeq {m\over k} - {m\over p_0} -N(1 - z_N^\theta) -(1-x_1){m\over p_0} \eqno (3.7) $$
which depends on the effective distance of the final momentum (energy) from the kinematical boundary for the $N$-fold process.
The Jacobian of the azimuthal variables transformation squared is
$$ J_N^2(z) = Det\, ||a_N||, \eqno (3.8)$$
where the matrix $||a_N||$ defines the quadratic form $Q_N(z,\varphi_k)$ which enters the argument of the $\delta$-function in Eq. $(3.6)$:
$$ Q_N(z,\varphi_k) = a_{kl} \varphi_k \varphi_l =\sum_{k=1}^{N} \varphi_k^2 -{\varphi_k\varphi_{k-1}\over z}. \eqno (3.9)$$
For example,
$$ Q_3(z,\varphi_k) =  \varphi_1^2 + \varphi_2^2 -\varphi_1 \varphi_2/z; \quad Q_4(z,\varphi_k)= \varphi_1^2 +\varphi_2^2 +\varphi_3^2 -
(\varphi_1\varphi_2 + \varphi_2\varphi_3)/z ,$$
$$ \quad Q_5(z,\varphi_k)= \varphi_1^2 +\varphi_2^2 +\varphi_3^2 +\varphi_4^2-
(\varphi_1\varphi_2 + \varphi_2\varphi_3 + \varphi_3\varphi_4)/z ,\eqno (3.9a) $$
see next section.

The phase space of the process in $(3.1)$ which depends strongly on $\Delta_N^{ext}$,  after integration over angular 
variables can be presented in the form
$$\Phi_N^{pions} = {1\over \omega_N'}\delta(m+\omega_{N-1}-\omega_N-\omega_N')\prod_{l=1}^Nd\Omega_l = {I_N(\Delta_N^{ext})\over k k_{N-1}}=
\frac{(\sqrt 2\pi)^{N-1}(\Delta_N^{ext})^{N-2}}{kk_{N-1}(N-2)!\sqrt N J_N(z_N^\theta)\left(z_N^\theta\right)^{N-1}} \eqno(3.10) $$

The normal Fermi motion of target nucleons inside of the nucleus increases the phase space considerably \cite{kop1, long}:
$$\Delta^{ext}_{N} = \Delta^{ext}_{N}|_{p_F=0} + \vec p^F_l\vec r_l/2m, \eqno (3.11) $$
where $\vec r_l =2m(\vec k_l-\vec k_{l-1})/k_lk_{l-1} $. A reasonable approximation is to take vectors $\vec r_l$ according to
the optimal kinematics for the whole process, and the Fermi momenta distribution of nucleons inside of the nucleus
in the form of the step function.
Integration over the Fermi motion leads to increase of the power of $\Delta_N^{ext}$ and change of numerical coefficients in
the expression for the phase space, details can be found in \cite{kop1, long}.

For the case of the nucleons rescattering there are some important differences from the light particle case, but
the quadratic form which enters the angular phase space of the process is essentially the same, with additional
coefficient:
$$\Phi_N^{nucleons} = 
{1\over k(m+\omega_{N-1})}\int \delta\left[\Delta^{ext}_{N,nucl} -\left(z_N^\theta\right)^N Q_N(\varphi_k) -{\left(z_N^\theta\right)^{N-2}\over 2}\sum_{l=1}^N\vartheta_l^2\right]
\prod_{l=1}^Nd\Omega_l =$$
$$=\left(\frac{\sqrt 2\pi}{\zeta_0 z^{N-1}}\right)^{N-1} \frac{(\Delta^{ext}_{N,nucl})^{N-2}}{(N-2)!\sqrt N J_N(z_N^\theta)}
\frac{(1-\zeta_N^2)(1-\zeta_{N-1}^2)}{4m^2\zeta_N} \eqno(3.12) $$
where
$$ \Delta^{ext}_{N,nucl} = \zeta_N-(1-x_1)\zeta_N{1-\zeta_1^2\over 1+\zeta_1^2} - {k\over m+\omega},  \eqno (3.13) $$
with $\zeta_N = \zeta_0 \left(z_N^\theta\right)^N,\; \zeta_1=\zeta_0z_N^\theta$.
As in the case of the light particle rescattering, the normal Fermi motion of nucleons inside the nucleus can be taken into account.

\section{The backward focusing effect (Buddha's light of cumulative particles)}
This is the sharp enhancement of the production cross section near the strictly 
backward direction, $\theta = \pi$. 
This effect has been noted first experimentally in Dubna (incident protons, final particles pions, protons and deuterons) \cite{baldin-77,foc-1}
and somewhat later by Leksin's group at ITEP (incident protons of 7.5 GeV/c, emitted protons
of 0.5 GeV/c) \cite{foc-2}.

In the papers \cite{kop1,long} where the small phase space method has been developed, 
it was noted that this effect can appear due to multiple interaction processes (see p.122 of \cite{long}). However, the
consideration of this effect was not detailed enough, the explicit angular dependence of the cross section near
backward direction, $\theta = \pi$, has not been established, estimates and comparison with data have not been made
\footnote{\small One of the authors (VBK) discussed the cumulative (backward) particles production off nuclei 
with professor Ya.A.Smorodinsky
who noted its analogy with known optical phenomenon - glory, or "Buddha's light". The
glory effect has been mentioned by Leksin and collaborators \cite{glo-1}, however,
it was not clear to authors of \cite{glo-1}, can it be related to cumulative production,
or not. In the case of the optical (atmospheric) glory phenomenon the light scatterings 
take place within droplets of
water, or another liquid. A variant of the atmospheric glory theory can be found in \cite{khare}. 
However, the existing explanation of the optical glory
is still incomplete, see, e.g. http://www.atoptics.co.uk/droplets/glofeat.htm. In nuclear physics
the glory-like phenomenon due to Coulomb interaction has been studied in \cite{maiorova} for the case of low energy
antiprotons (energy up to few KeV) interacting with heavy nuclei.}. 

The backward focusing effect has been observed and confirmed later in a number of
papers for different projectiles and incident energies \cite{foc-4,glo-1,glo-2}. 
It seems to be difficult to explain the backward focusing effect as coming from 
interaction with dense few nucleon clusters existing inside the nucleus.

Mathematically the focusing effect comes from the consideration of the small phase space of 
the whole multiple interaction process by the method described in previous section and in \cite{kop1,long}.
It takes place for any MIP, regardless the particular
kind of particles or resonances in the intermediate states.
As it was explained in section 2, when the angle of cumulative particle emission is large, but different from $\theta = \pi$,
there is a prefered plane for the whole process.
When the final angle $\theta = \pi$, then integration over one of azimuthal angles takes place
for the whole interval $[0, 2\pi]$, which leads to the rapid increase of the resulting cross section when the
final angle $\theta$ approaches $\pi$. 

We show first that the azimuthal (axial) focusing takes place for any values of the polar scattering angles $\theta_k^{opt}$.
For arbitrary angles $\theta_k$ the cosine of the angle between directions $\vec n_k$ and $\vec n_{k-1}$ is
$$z_k= (\vec n_k \vec n_{k-1}) \simeq cos (\theta_k - \theta_{k- 1}) (1-\vartheta_k^2/2) -sin (\theta_k - \theta_{k- 1}) \vartheta_k 
+ sin (\theta_k) sin \theta_{k-1} (\varphi_k -\varphi_{k-1})^2/2. \eqno (4.1)$$

After substitution $sin \theta_k\varphi_k \to \varphi_k$ we obtain
$$z_k= (\vec n_k \vec n_{k-1}) \simeq cos (\theta_k - \theta_{k- 1}) (1-\vartheta_k^2/2) -sin \left(\theta_k - \theta_{k- 1}\right) \vartheta_k 
+ {s_{k-1}\over 2s_k}\varphi_k^2 + {s_{k}\over 2s_{k-1}}\varphi_{k-1}^2-\varphi_{k-1}\varphi_k, \eqno (4.2)$$
where we introduced shorter notations $s_k = sin \theta_k$.

It follows from Eq. $(4.2)$ that in general case of arbitrary polar angles $\theta_k$ the quadratic form depending on the 
small azimuthal deviations $\varphi_k$ which enters the sum $\sum_k (1-z_k)$ for the $N$-fold process is
$$Q_N^{gen}(\varphi_k,\varphi_l) = {s_2\over s_1}\varphi_1^2 + {s_1+s_3\over s_2}\varphi_2^2 +{s_2+s_4\over s_3}\varphi_3^2 +....
+ {s_{N-2}+s_N\over s_{N-1}}\varphi_{N-1}^2 -$$
$$-2 \varphi_1\varphi_2 -2\varphi_2\varphi_3 - ... - 2\varphi_{N-2}\varphi_{N-1} =||a||^{gen}(\theta_1,...,\theta_{N-1})_{kl}
\varphi_k\varphi_l , \eqno(4.3)$$
with $s_N= sin \theta$.
E.g., for $N=6$ we have the matrix
$$||a||_{N=6}^{gen}(\theta_1,\theta_2,\theta_3,\theta_4, \theta_5)=
\left[\begin {array}{ccccc} s_{2}/s_{1} & -1 & 0 & 0 & 0\\ 
-1 & (s_{1}+s_{3})/s_{2} & -1  & 0 & 0  \\
 0 & -1  & (s_{2}+s_{4})/s_{3}& -1 & 0\\
0  &  0 & -1 & (s_{3}+s_{5})/s_{4}&-1\\
0  &  0 & 0 & -1 & (s_{4}+s_{\theta})/s_{5}
   \end {array}\right], \eqno (4.4) $$
$s_\theta = s_6$, and generalization to arbitrary $N$ is straightforward.

Determinant of this matrix can be easily calculated.
It can be shown by induction that at arbitrary $N$ (for details see \cite{kmp})
$$ Det \left(||a||_N^{gen}\right) = {s_\theta\over s_1}, \quad s_\theta = s_N. \eqno(4.5) $$

After integration the delta-function containing the quadratic form over the small azimuthal deviations we obtain
$$ \int \delta \left(\Delta - ||a||^{gen}_N(\theta_1,...,\theta_{N-1})_{kl}\varphi_k\varphi_l \right) d\varphi_1... d\varphi_{N-1} 
= {\Delta^{(N-3)/2}\over Det ||a||_N^{gen} (N-3)!!} (2\pi)^{(N-3)/2}c_{N-3}= $$
$$=\sqrt{{s_1\over s_\theta}} {\Delta^{(N-3)/2}\over  (N-3)!!} (2\pi)^{(N-3)/2}c_{N-3} , \eqno(4.6) $$
$c_n=\pi$ for odd $n$, and $c_n=\sqrt{2\pi}$ for even $n$, and $N-3\geq 0$.

We obtain from above expressions the characteristic angular dependence of the cumulative particles production cross section  near $\theta =\pi$:
$$ d\sigma \sim \sqrt{{s_1\over s_\theta}} \simeq \sqrt{{s_1\over \pi-\theta}}, \eqno(4.7) $$
since $sin\theta \simeq \pi -\theta$ for $\pi-\theta \ll 1$.

This formula does not work at $\theta = \pi$, because
integration over the azimuthal angle which defines the plane of the whole MIP takes place in the interval $(0, 2\pi)$. The result for the 
cross section is final, of course, as we have shown in details for the case of the optimal kinematics \cite{kmp}.
In this case the equality of the polar scattering angles takes place, $\theta_k= k\theta/N$ (see section 2), and the general quadratic form 
goes over into quadratic form obtained in \cite{long} with some coefficiens:
$$ Q^{gen} \to  2 z_N^\theta Q(z_N^\theta, \varphi_k,\varphi_l), \quad z_N^\theta = cos(\theta/N), \eqno (4.8) $$
and
$$Det (||a||_N^{gen}) =\left(2 z^\theta_N\right)^{N-1} Det (||a||_N).  \eqno (4.8a) $$
It is convenient to present the quadratic form which enters the $\delta$ - function in $(3.6)$ as
$$ Q_N(z_N^\theta,\varphi_k,\varphi_l) =J_2^2\left(\varphi_1-{\varphi_2\over 2zJ_2^2}\right)^2 +{J_3^2\over J^2_2}
\left(\varphi_2-{J_2^2\varphi_3\over 2zJ_3^2}\right)^2 +...$$
$$...  + {J_{N-1}^2\over J_{N-2}^2}\left(\varphi_{N-2}-{J_{N-2}^2\varphi_{N-1}\over 2z J_{N-1}^2}\right)^2 +
{J_N^2\over J_{N-1}^2} \varphi_{N-1}^2 .\eqno (4.9)$$
For the sake of brevity we omitted here the dependence of all $J_k^2$ on their common argument $z_N^\theta $.  The recurrent relation 
$$J_N^2(z)= J_{N-1}^2(z)-{1\over 4z^2}J_{N-2}^2(z) \eqno (4.10) $$ 
can be obtained from $(4.9)$, since, as it follows from$(3.6)$ and $(3.9)$
$$Q_{N+1}(z,\varphi_k,\varphi_l)  = Q_N(z, \varphi_k,\varphi_l) +\varphi_N^2 -\varphi_N\,\varphi_{N-l}/z \eqno (4.11)$$
(recall that for the $N+1$-fold process $\varphi_{N+1} =0$ by definition of the whole plane of the process),

The following formula for $J_N^2(z_N^\theta)$ has been obtained in \cite{long}:
$$Det ||a_{kl}|| =J_N^2(z_N^\theta) = 1 + \sum_{m=1}^{m < N/2}\left(-{1\over 4\left(z_N^\theta\right)^2}\right)^m 
{\prod_{k=1}^{m}(N-m-k)\over m!}. \eqno (4.12)  $$
Recurrent relations for Jacobians with subsequent values of $N$ and with same argument $z$:
$$ J^2_{N+1}(z) = J^2_N(z) -{1\over 4z^2} J^2_{N-1}(z)= 
J^2_{N-1}(z)\left(1-{1\over 4z^2}\right) -{1\over 4z^2} J^2_{N-2}(z)\eqno (4.13) $$
can be continued easily to lower values of $N$ and also used for calculations of $J_N^2$ at any $N$ starting
from two known values, $J_2^2(z) =1$ and $J_3^2(z) =1-1/(4z^2)$  (see \cite{kmp}).
The Eq. $(4.12)$ can be confirmed in this way.

The condition $J_N(\pi/N) =0$ leads to the equation for $z_N^\pi$ which solution
(one of all possible roots) provides the value of $cos(\pi/N)$ in terms of radicals.
\footnote{In other words, in \cite{long,kmp} we found a way to get polinomials in $1/z^2$ with rational coefficients, one
of roots of which is just $cos(\pi/N).$}
The following expressions for these Jacobians take place \cite{kop1,long,kmp}
$$J_2^2(z)=1; \qquad J_3^2(z) = 1-{1\over 4z^2}; \qquad J_4^2(z)= 1-{1\over 2z^2}, \eqno (4.14) $$
$J_3(\pi/3)=J_3(z=1/2)=0$,   $J_4(\pi/4)=J_4(z=1/\sqrt 2)=0$.
 For $N=5$ 
$$J_5^2=1-{3\over 4z^2}+{1\over 16 z^4}, \qquad  (J_5^2)'_z = {3\over 2z^3} - {1 \over 4 z^5} \eqno (4.15) $$
and one obtains $cos^2(\pi/5) = (3+\sqrt 5)/8$, $J_5(\pi/5)=0$.

At $N=6$
$$J_6^2=1-{1\over z^2}+{3\over 16 z^4} = J_3^2 \left(1-{3\over 4z^2}\right), \qquad (J_6^2)'_z = {2\over z^3} - {3\over 4 z^5}. \eqno (4.16)$$
For $N=7$
$$J_7^2=1-{5\over 4z^2}+{3\over 8z^4} -{1\over 64z^6}, \qquad  (J_7^2)'_z = {5\over 2 z^3} - {3\over 2 z^5} + {3\over 32 z^7}, \qquad J_7(\pi/7)=0. \eqno (4.17)$$
$$J_8^2=1-{3\over 2z^2}+{5\over 8z^4} -{1\over 16z^6}=J_4^2 \left(1-{1\over z^2}+{1\over 8z^4}\right), 
\qquad (J_8^2)'_z = {3\over z^3} - {5\over 2 z^5} +{3\over 8 z^7}, \qquad J_8(\pi/8)=0  \eqno (4.18)$$
For arbitrary $N$,  $J_N^2$ is a polinomial in $1/4z^2$ of the power $|(N-1)/2|$ (integer part of $(N-1)/2$), see Eq. $(4.14)$.
These equations can be obtained using the elementary mathematics methods as well, see \cite{kmp}, Appendix.
The case $N=2$ is a special one, because $J_2(z)=1$ - is a constant. In this case the 2-fold
process at $\theta =\pi$ (strictly backwards) has no advantage in comparison with the direct one, see Eq. $(2.5)$,
if we consider the elastic rescatterings.

For particles emitted strictly backwards the phase space has different form, instead of
$J_N(\theta/N)$ enters $J_{N-1}(\theta/N)$ which is different from zero at $\theta =\pi$, and we have instead of Eq. (3.6)
$$ I_N(\varphi, \vartheta) = \int \delta\biggl[\Delta_N^{ext} - z_N^\pi\bigg(\sum_{k=1}^{N} \varphi_k^2 -\varphi_k\varphi_{k-1}/z_N^\pi
+\vartheta_k^2/2\biggr)\biggr] \left[\prod_{l=1}^{N-2} d\varphi_ld\vartheta_l\right] 2\pi d\vartheta_{N-1} = $$
$$= \frac{\left(\Delta_N^{ext}\right)^{N-5/2} (2\sqrt 2 \pi)^{N-1}}{J_{N-1}(z_N^\pi) \sqrt N (2N-5)!! \left(z_N^\pi\right)^{N-3/2}}, \eqno (4.19) $$
This follows from Eq. $(4.11)$ where at $\theta = \pi$ the last term disappears, since $J_N(\pi/N) = 0$ and integration over
$d \varphi_{N-1}$ takes place over the whole $2\pi$ interval.

\begin{figure}[h]
\begin{center}
\includegraphics[natwidth=800,natheight=600,width=15.cm,angle=0]{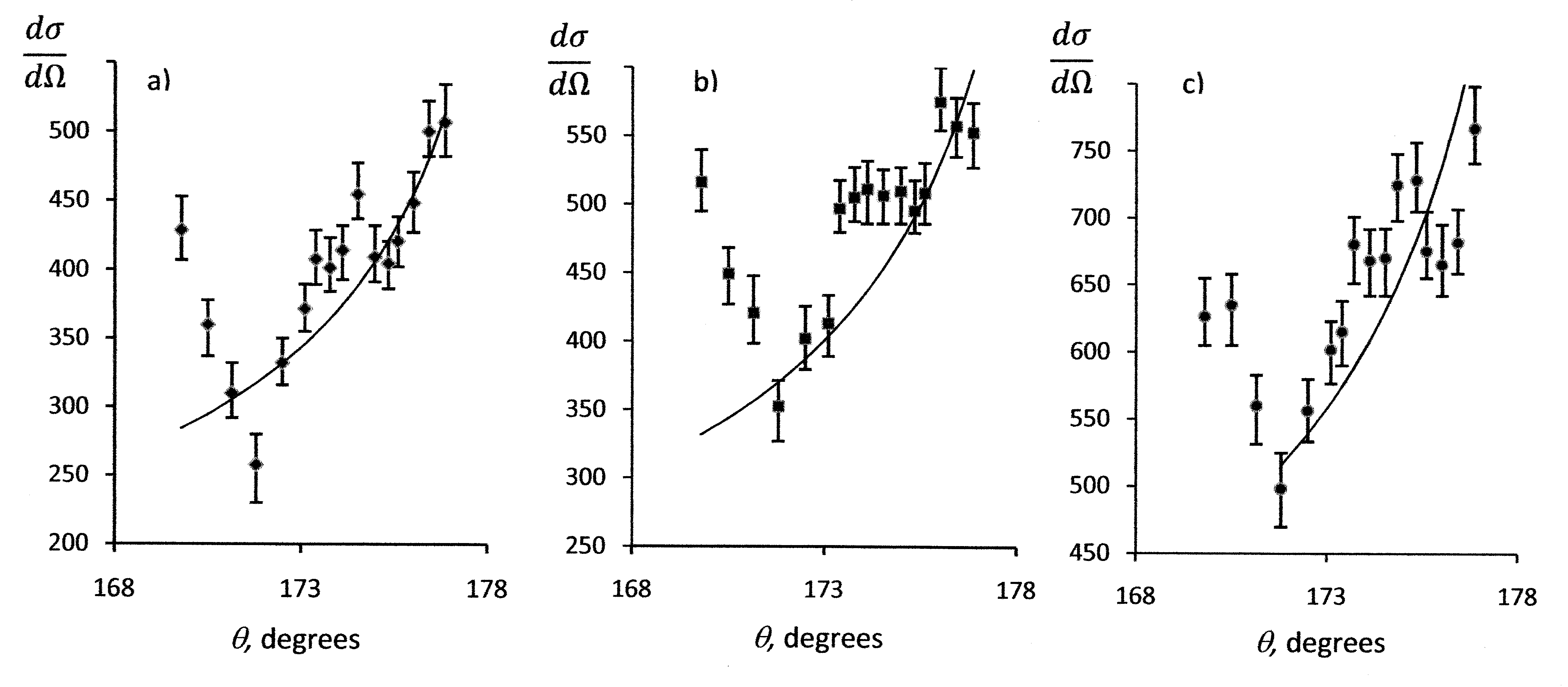}
\end{center}
{\baselineskip=9pt {\small   {\bf Fig.\ 4.1.} Angular distributions of secondary protons with kinetic energy between $0.11$ and $0.24\, GeV$ 
emitted from the Pb nucleus, in arbitrary units. The momentum of the projectile protons is $4.5\, GeV/c$. 
 a) The energy of emitted protons in the interval $0.11 \, - \, 0.24\, GeV$; b) the energy interval $0.08\,- \, 0.11\, GeV$; 
c) the energy interval  $0.06\,-\, 0.08\, GeV$.  Data obtained by G.A.Leksin group at ITEP, taken 
from Fig. 3 of paper \cite{glo-1}.} The curves are drawn according to the formula $A + B/\sqrt{\pi - \theta}$, where $A$ and $B$ are some fitted
constants.}
\end{figure}
\vglue 0.1cm

To illustrate the azimuthal, or axial focusing which takes place near $\theta = \pi$ the ratio is useful of 
the phase spaces near the backward direction and  strictly at $\theta =\pi$. The ratio of the observed cross sections in the interval
of several degrees slightly depends on the elementary cross sections and is defined mainly by this ratio of phase spaces.
It is
$$ R_N(\theta) = {\Phi(z)\over \Phi (\theta =\pi)} = \sqrt{\Delta_N^{ext}\over z_N^\pi}  {(2n-5)!! \over 2^{N-1} (N-2)!}{J_{N-1}(z_N^\pi)\over sin(\pi/N) J_N(z_N^\theta)} 
\eqno (4.20)  $$
Near $\theta = \pi$ we use that 
$$J_N(z_N^\theta) \simeq \sqrt{{\pi - \theta\over N} [J_N^2]'(z_N^\pi) sin{\pi\over N} } \eqno (4.21) $$
and thus we get
$$ R_N(\theta) = C_N \sqrt{{\Delta_N^{ext}\over \pi - \theta}} \eqno (4.22) $$
with
$$ C_N = {J_{N-1}(z_N^\pi) \sqrt N \over [(J_N^2)'(z_N^\pi)]^{1/2} [sin(\pi/N)]^{3/2}} { (2N-5)!! \over \sqrt{z_N^\pi} (N-2)! 2^{N-1}} \eqno(4.23)  $$
We need also values of $J_{N-1}[\pi/N] $ to estimate the behaviour of the cross sction near $\theta =\pi$.
Integration over variable $x_1$ leads to multiplication $C_N$ by factor $(2N-3)/(2N-2)$, i.e. it makes it smaller, increasing the
effect under consideration.
The constant $C_N$ variates between $0.38$ and $0.26$ for $N$ between $3$ and $7$, slightly decreasing with increasing $N$ \cite{kmp}.

Inclusion of resonance excitation in one (or several) intermediate states leads to the increase of the quantity $\Delta^{ext}_N$
according to formulas of section 2, and to the increase of the phase space of the whole MIP, but the effect of azimuthal focusing
persists.
Quite similar results can be obtained for the case of nucleons, only some technical details are different, see \cite{kmp}.
The inclusion of the normal Fermi motion of nucleons inside the nucleus increases the
values of $\Delta_N^{ext}$, but numerical coefficient in $C_N$ becomes smaller.
The behaviour given by Eq. (4.15) is in good agreement with available data, the value of the constants $C_N$ is not
important for our semiquantiatative treatment.

There are other data besides shown at Fig. 4.1 (incident protons of $4.5\,GeV/c$, detected cumulative particle also proton),
where the agreement of the $1/\sqrt{(\pi - \theta)}$ law with data is quite good, see \cite{kmp}. The flat behaviour of 
the differential cross section near the backward direction has been observed in several experiments, in particular, at the Yerevan Physical Institute.

\section{Discussion and conclusions}
The nature of the cumulative particles is complicated and not well understood so far.
There are different possible sources of their origin, one of them are the multiple 
collisions inside the nucleus, i.e. elastic or inelastic rescatterings.
We have shown that the enhancement of the particles production cross section off nuclei near
the backward direction, the glory-like backward focusing effect, is a natural property
of the multiple interaction mechanism for the cumulative particles production.
It takes place for any multiplicity of the process, $N\geq 3$, when the momentum of the emitted particle is close 
to the corresponding kinematical boundary. The universal dependence of the cross section,
$d\sigma \sim 1/\sqrt{\pi - \theta}$ near the final angle $\theta \sim \pi$, takes place regardless 
the multiplicity of the process. This statement by itself is quite rigorous and presented for the first time in the literature. 
The competition of the
processes of different multiplicities can make this effect difficult for observation in some cases.
Presently we can speak only about qualitative,
in some cases semiquantitative agreement with data. It is not clear yet how the transition to the
strictly backward direction proceeds. The angular distribution of emitted
particles near $\theta = \pi$ can have a narrow dip, i.e. it may be of a crater (funnel)-like form. Further studies, analytical
and, probably, numerical as well, are necessary to clear up this point. It is worse noting here that
measurements of the differential cross section at $\theta = \pi$ are difficult and even not possible in experiments
with the target nucleus at rest.

This effect, observed in a number of experiments at JINR and ITEP, is a clear manifestions 
of the fact that multiple interactions make important contribution
to the cumulative particles production cross sections. However, this observation does not exclude substantial contribution
of interaction of the projectile with the few-nucleon, or multiquark clusters possibly existing in nuclei.
We have proved rigorously the existence of the azimuthal focusing for arbitrary polar angles in case of the light particles
rescattering, and for the 
case of the optimal (basic) configuration of the MIP, also for nucleons rescattering. Oobviously, the azimuthal
focusing, discussed e.g. in \cite{khare} for the optical glory phenomenon, takes place for any kind of
MIP, only some technical details are different.

It would be important to detect the backward focusing effect for different types of produced
particles, including hyperons and kaons. This effect can be considered as a "smoking gun" of the MIP mechanism. 
If this nuclear glory-like phenomenon is observed for all kinds
of cumulative particles, its universality would be a strong argument in favor of importance of MIP.
The role of the multiple interaction processes leading to the large
angle particles production off nuclei is certainly underestimated, still, by many authors.
Possible cosmophysical consequences of this effect may be of interest.
Further efforts are necessary to settle this difficult and important 
challenge of disentangling between the nontrivial effects of the nuclear structure and the MIP contributions.\\

We are indebted to V.M.Lobashev who supported strongly 
the main idea that the background
multiple interaction processes should be investigated and their contribution
should be subtracted from measured cross sections to determine the weight of few-nucleon or 
multiquark clusters in nuclei.
We thank S.S.Shimansky, 
B.Z.Kopeliovich, A.P.Krutenkova, A.B.Kurepin, V.L.Matushko for useful remarks and discussions.

This work is supported in part by Fondecyt (Chile), grant number 1130549.\\

\end{document}